\documentclass[pre,amsmath,amssymb,twocolumn,superscriptaddress]{revtex4-2}
\usepackage{amsmath,amssymb}
\usepackage[usenames]{color}
\usepackage{amssymb}
\usepackage{dsfont}
\usepackage{grffile}
\usepackage{textcomp}
\usepackage{xspace}
\usepackage{hyperref}
\usepackage{graphicx} %
\usepackage{outlines}
\usepackage{natbib}
\usepackage[dvipsnames]{xcolor} 
\usepackage[export]{adjustbox}
\usepackage{ulem} 
\usepackage{tabularx} 

\newcommand{\tr}{{\rm tr}}

\newcommand{\im}{{\rm i}}

\newcommand{\PCA}{\mathcal{V}} 
\DeclareMathOperator{\Tr}{Tr}
\newcommand{\dt}{\Delta t} 

\newcommand{\sutdphys}{Science, Mathematics and Technology Cluster, Singapore
University of Technology and Design, 8 Somapah Road, 487372 Singapore}
\newcommand{\ictp}{The Abdus Salam International Centre for Theoretical Physics, Strada Costiera 11, 34151 Trieste, Italy}
\newcommand{\sutdepd}{EPD Pillar, Singapore University of Technology and Design, 8 Somapah Road, 487372 Singapore}
\newcommand{\ihpc}{Institute of High-Performance Computing, Agency for Science, Technology, and Research (A*STAR), Singapore}

\begin{document}
\title{Complexity of spin configurations dynamics due to unitary evolution and periodic projective measurements}

\author{Heitor P. Casagrande}
\email{heitor\_peres@mymail.sutd.edu.sg} 
\affiliation{\sutdphys}

\author{Bo Xing}
\affiliation{\sutdphys}

\author{Marcello Dalmonte}
\affiliation{\ictp}

\author{Alex Rodriguez}
\affiliation{\ictp}

\author{Vinitha Balachandran}
\affiliation{\sutdphys}
\affiliation{\ihpc}

\author{Dario Poletti}
\email{dario\_poletti@sutd.edu.sg} 
\affiliation{\sutdphys}
\affiliation{\sutdepd}
\affiliation{\ictp}

\begin{abstract}
We study the Hamiltonian dynamics of a many-body quantum system subjected to periodic projective measurements which leads to probabilistic cellular automata dynamics.
Given a sequence of measured values, we characterize their dynamics by performing a principal component analysis. 
The number of principal components required for an almost complete description of the system, which is a measure of complexity we refer to as PCA complexity, is studied as a function of the Hamiltonian parameters and measurement intervals. 
We consider different Hamiltonians that describe interacting, non-interacting, integrable, and non-integrable systems, including random local Hamiltonians and translational invariant random local Hamiltonians. 
In all these scenarios, we find that the PCA complexity grows rapidly in time before approaching a plateau. 
The dynamics of the PCA complexity can vary quantitatively and qualitatively as a function of the Hamiltonian parameters and measurement protocol. 
Importantly, the dynamics of PCA complexity present behavior that is considerably less sensitive to the specific system parameters for models which lack simple local dynamics, as is often the case in non-integrable models. In particular, we point out a figure of merit that considers the local dynamics and the measurement direction to predict the sensitivity of the PCA complexity dynamics to the system parameters. 
\end{abstract}
  
\maketitle

\setcounter{figure}{0}


\section{Introduction} 
One of the most intriguing aspects of quantum mechanics is the role of the measurement \cite{DiracBook}, which has motivated numerous theoretical, experimental, and philosophical investigations.
Projective measurements result in the need to update the description of the system (sometimes referred to as the collapse of the wave function) and weaker measurements can lead to less significant feedback on the quantum system \cite{WisemanMilburn2010}. 
In recent years, it was shown that quantum measurements can significantly affect quantum thermodynamics processes \cite{TalknerHanggi2016, ZhengPoletti2016, CampisiHanggi2011}, and can be used as a thermodynamics source akin to a heat bath \cite{JordanAuffeves2020, DingTalkner2018}.
In a very different research direction, much attention has been given to understanding measurement-induced phase transitions \cite{LiFisher2018, LiFisher2019, SkinnerNahum, ChanSmith2019, TurkeshiFazio2021, NahumRuhman2021, AgrawalVasseur2022, NoelMonroe2022, weinstein2022scrambling, kelly2022coherence, PhysRevApplied.14.024054, ChiriacoDalmonte2023}. In this case, following the quantum trajectories stemming from a measurement protocol and studying the entanglement entropy in the corresponding quantum states, one can observe a phase transition between area-law and volume-law entanglements depending on the measurement rate. 
Measuring a portion of a system can also give insight into the thermalization dynamics of the non-measured system, because of the correlations they have with the measured portion.
This leads to the study of the projected ensemble and deep thermalization \cite{HoChoi2022, IppolitiHo2022, CotlerChoi2023}.

In this work, we consider the dynamics of a many-body quantum system that undergoes Hamiltonian evolution.
At regular intervals, each site of the system is subjected to projective measurements.
After each set of measurements, the system returns to a single-site product state and continues to evolve unitarily until the next projective measurement.
For concreteness, we consider measurements on a spin chain that detect locally if each spin is pointing up or down in a particular direction. 
This measurement protocol results in the generation of a sequence of (classical) spin configurations which can be thought of as a probabilistic cellular automaton \cite{LouisNardi2018}.
We study these sequences as a function of the type of unitary dynamics and measurement frequency.
From the sequence of configurations, we devise a correlation matrix that stores the correlation between different sites averaged over time, and perform a principal component analysis (PCA) on it. This returns the number of principal components required to have almost complete description of the system, and we refer to it as PCA complexity.
Note that PCA has already been used to characterize the emergence of different phases of matter \cite{Wang2016, HuScalettar2017, WangZhai2017, ThiagoMendesRodriguez2021, Wetzel2017, CostaRajiv2017, KhatamiScalettar2020, Turkeshi2022}; alternatively, \cite{CarleoZdeborova2019, Carrasquilla2020} are reviews of works which use machine learning tools to characterize matter. 
For the unitary evolution, we consider different classes of Hamiltonians, from non-interacting to non-integrable,  disordered to homogeneous. 
We observe that the principal components grow rapidly in time and reach a plateau. 
Quantitatively, the evolution of the PCA complexity is much less dependent on the details of the Hamiltonian terms for non-integrable interacting systems. 
We further find that the evolution is strongly influenced by the pinning coefficient, which indicates the relative weight of the Hamiltonian terms which retain the configuration that has been chosen by the projective measurements. 

This paper is structured as follows: we introduce the projective measurement protocol, the different Hamiltonians, and the quantities we use to describe the dynamics in Sec.~\ref{sec:protocol}. In Sec.~\ref{sec:results}, we compile our results and finally, we draw conclusions in Sec.~\ref{sec:conclusions}.

\section{Protocol, model and quantities studied} \label{sec:protocol}

In this section, we describe the projective measurement protocol used to generate the probabilistic cellular automaton.  
The measurement protocol consists of four steps:
\begin{enumerate}
    \item A random initial state is prepared by taking a product state with no superpositions in the computational basis, i.e. a state of the type $|\!\uparrow\downarrow\downarrow\uparrow\uparrow\downarrow\rangle$. 
    \item This initial state is subject to unitary evolution via the Hamiltonian $H$ for a time $\Delta t$. 
    \item Projective measurements are performed and the system is brought back to a new product state.

    \item Steps 2. and 3. are repeated for $m$ times to generate $m$ configurations.
\end{enumerate}

\begin{figure}[h]
\includegraphics[width = \columnwidth]{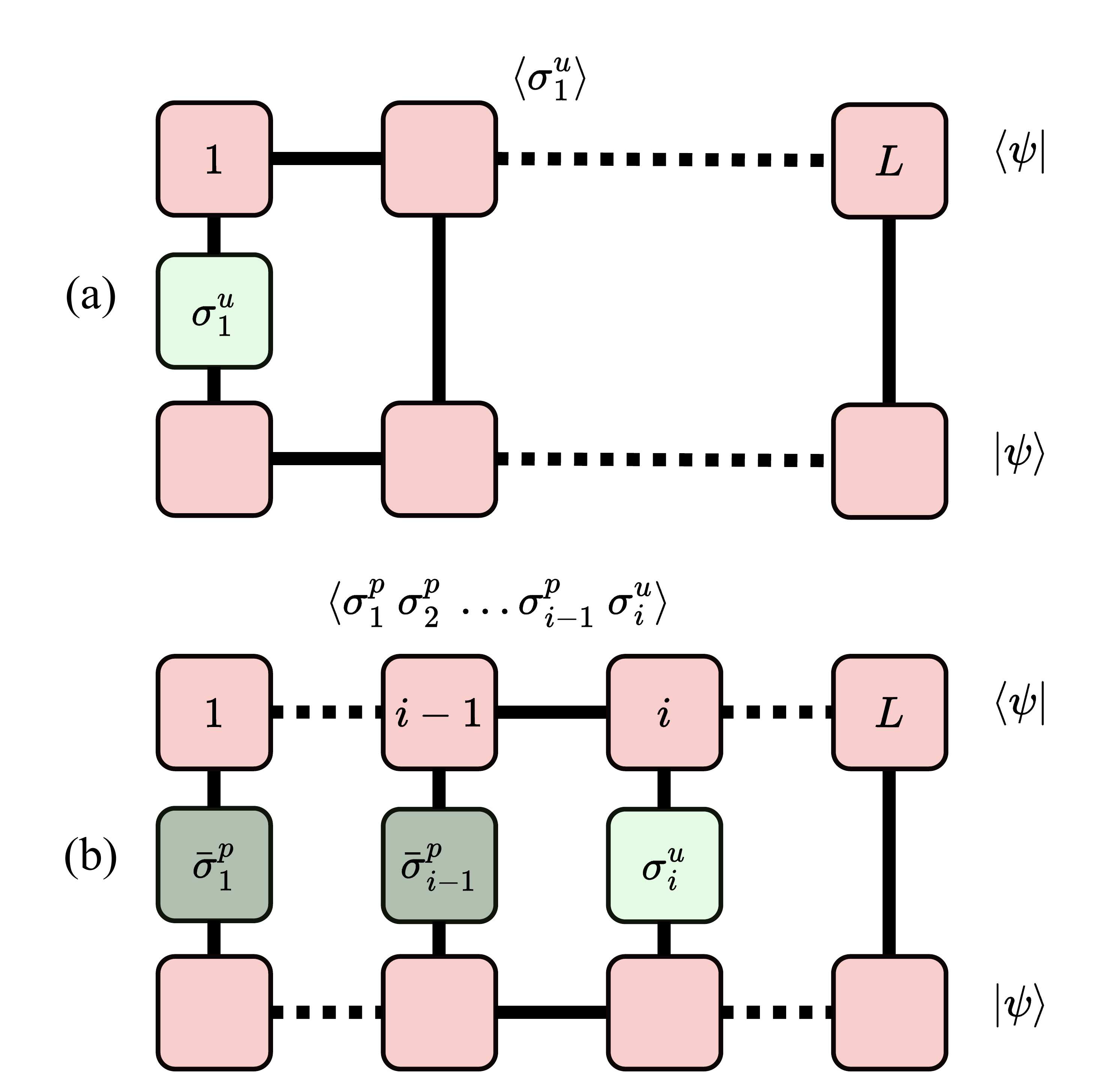}
	    \caption{(a) Measuring $\sigma^u_1$ at the first site of the matrix product state representation of the system. (b) Measuring the conditional expected value of $\sigma_i^u$ given the first $i-1$ spins fixed in either up/down position depending on the result of the previous measurements.}
        \label{fig:zipperSetup}
\end{figure}

{\it Sampling from projective measurements protocol}:  
Given a wave function $|\psi\rangle$ we aim to sample a configuration drawn from the probability distribution $|\langle \sigma_1 \sigma_2 \dots \sigma_L  | \psi
 \rangle|^2$, where each $\sigma_i$ can be in the $\uparrow$ or $\downarrow$ configuration. 
To extract a configuration one can use the Metropolis-Hastings algorithm \cite{MetropolisTeller1953, Hastings1970}. 
In this work, we use a zipper approach, as explained in \cite{HanZhang2018}. The advantage of this approach is that the sampling is direct, with no autocorrelation between subsequent samples (this is particularly important to ensure our analysis below is not influenced by spurious correlations). 
The zipper approach is particularly intuitive when the (exact) wavefunction is written as a matrix product state, and it consists in extracting local configurations from consecutive samplings over conditional probabilities, as depicted in Fig.\ref{fig:zipperSetup}: First, we measure the probability of the first spin being in the $\uparrow$ state, i.e., $P^\uparrow_1=\langle \psi| \sigma^u_1 |\psi\rangle$, where $\sigma^u_i=(1+\sigma^z_i)/2$ while $\sigma^d_i=(1-\sigma^z_i)/2$. 
Then, we draw a uniformly distributed random number $r_1$ between $0$ and $1$.
If $r_1 < P^\uparrow_1$, we consider the first spin to be pointing up. 
Otherwise, it is pointing down. 
We refer to this determined direction as $\bar{\sigma}^p_i$, where $p$ can be $u$ (up) or $d$ (down).
Following this approach, in order to evaluate the projected state of the $i-$th spin, we act in two steps.
First, we compute the conditional probability of the $i-$th spin being in the $\uparrow$ state, given the projected configuration of all previous spins, i.e., 
\begin{align}
     P^{\uparrow|\bar{\sigma}^p_{i-1}\dots\bar{\sigma}^p_{1}}_i= \frac{ \langle \psi| \sigma^u_i \bar{\sigma}^p_{i-1}\dots\bar{\sigma}^p_{1} |\psi\rangle}{\langle \psi| \bar{\sigma}^p_{i-1}\dots\bar{\sigma}^p_{1} |\psi\rangle}. 
\end{align}
Then, we draw a uniformly distributed random number $r_i$ between $0$ and $1$.
If $r_i < P^{\uparrow|\bar{\sigma}^p_{i-1}\dots\bar{\sigma}^p_{1}}_i$, the spin is in the $\uparrow$ state.
Otherwise, it is in the $\downarrow$ state.

{\it Models}: To study qualitatively different systems, we consider the following generic nearest neighbor Hamiltonian in one dimension
\begin{align}
    H & = \sum_{i=1}^{L-1} \left( J_{x,i} \sigma^{x}_{i} \sigma^{x}_{i+1} + J_{y,i} \sigma^{y}_{i} \sigma^{y}_{i+1} + J_{z,i} \sigma^{z}_{i} \sigma^{z}_{i+1} \right) \nonumber \\ 
  & + \sum_{i=1}^{L} \left( h_{x,i} \sigma^{x}_{i}  + h_{y,i} \sigma^{y}_{i} + h_{z,i} \sigma^{z}_{i} \right),  
\label{Hamiltonian}
\end{align}
where the coefficients $J_{a,i}$ and $h_{a,i}$, with $a=x,y,z$, can be different on different sites. Henceforth we work in units such that one of the terms, typically $J_{z,i}$ is equal to one, and so is $\hbar$. 
We focus mainly on one integrable and non-interacting case (Ising model $H_I$)\footnote{Non-interacting in the sense that after Jordan-Wigner transformation \cite{JW} it reduces to a free fermions Hamiltonian.}, one Integrable and Interacting case (XYZ model $H_{XYZ}$) and one non-integrable and also interacting case (XXZ model with transverse and longitudinal fields $H_H$). Additionally, since we acknowledge that the chosen measurement axis plays an important role in the dynamics, we also study the rotated version of such models, changing the tunnelings and the interactions around. We refer to these Hamiltonians as ``rotated'' ones and use, respectively, the symbols $H_{IR}$, $H_{XYZR}$, and $H_{HR}$. 
Additionally, we also consider random Hamiltonians with either all-to-all or local couplings. 
For local couplings, we simply consider the Hamiltonian in Eq.~(\ref{Hamiltonian}) using random coefficients $J_{a,i}$ and $h_{a,i}$. We refer to this Hamiltonian as $H_R$. For the case in which the coefficients $J_{a,i}$ and $h_{a,i}$ are random and translation-invariant, i.e., $J_{a,i}=J_a$ and $h_{a,i}=h_a$, we refer to the Hamiltonian as $H_{RT}$. 
Instead, for all-to-all couplings, we consider two types of Hamiltonians one that follows Wigner-Dyson statistics, and one that follows Poisson.  
For the first case we generate a random global Hamiltonian $H_{GG}$ which contains a diagonal drawn from $\mathcal{N} (0, \sqrt{2})$ and all other elements drawn from $\mathcal{N} (0, 1)$ \cite{Heiss2014}. 
For the second case, we first construct a diagonal matrix with Poisson statistics $H_{DP}$, constructed drawing random numbers from a normal distribution, and then we apply a unitary transformation $e^{-\im H_{GG}\Delta t}$ on it to give an all-to-all random matrix which follows Poisson level spacing statistics $H_{GP}$, i.e. 
\begin{align}
    H_{GP} = e^{-\im H_{GG}\Delta t} \; H_{DP} \; e^{\im H_{GG}\Delta t},   \label{eq:HGP} 
.\end{align}
 with $\Delta t=1$. 



{\it Monitored quantities}: After the $m-$th measurement, the spin at position $i$ can be in a configuration which we denote as either $z^m_i = 0$ or $1$, depending on whether the spin is down or up.
Therefore, a system of size $L$ will give a sequence of $L$ digits after every measurement.
In order to  understand the dynamics of these measured outcomes, we consider $N$ different trajectories of the same Hamiltonian at the same time. 
Hence, the data we collect after the $m$-th measurement is $\vec{z}^{\;m} = (z_1^m,\;z_2^m, \dots\; z_{LN}^m)$, which has a size of $L\times N$.
After $m$ measurements, the total size of data collected is $m$ rows of $L \times N$ columns. A set of $N$ trajectories will be referred to as a \textit{realization}.

The first quantity we use to study the dynamics of these probabilistic cellular automata is the Hamming distance $d_H$, given by 
\begin{align}
    d_H(m,n) = \dfrac{1}{N} \sum_{i=1}^{LN} z^m_i \oplus z_{i}^{n}, 
\end{align}
where $\oplus$ is the sum in $Z_2$, which gives $1$ when $z_i^m \neq z_i^n$ and $0$ otherwise. 

The other quantity we study is the principal components of the correlation matrix obtained from the evolution plus measurements protocol. 
More specifically, at a given time step $m$ of the evolution, we build the correlation matrix $C$ with elements
\begin{align}
    C_{i,j}^m = \sum_{n = 1}^{m} \dfrac{z_i^n z_j^n }{m} - \left( \sum_{n = 1}^{m} \dfrac{z_i^n}{m} \right) \left( \sum_{n = 1}^{m} \dfrac{z_j^n}{m} \right). 
    \label{eq:Cij} 
\end{align} 
From $C_{ij}^m$, we compute the eigenvalues $\lambda_l$, order them in decreasing manner, and consider the smallest number of eigenvalues $s$ which constitutes $95\%$ of the trace in magnitude.
The PCA complexity is then defined as
\begin{align}
    \PCA_m = \inf p \;\;\;{\rm such\;that }\;\;\frac{\sum_{l=1}^p \lambda_l}{\tr[C]}=0.95. 
\end{align} 
At the initial step $m=1$, the correlation matrix is $0$ everywhere.
If the dynamics are highly mixing, the correlation matrix will tend toward the identity matrix as $m$ increases.
In such a case, the maximum PCA complexity is obtained with $\PCA_m = 0.95LN$.

One way to characterize the Hamiltonians we consider 
is by evaluating their level spacing statistics, i.e., the ratio between consecutive eigenvalues $e_{n-1}, e_n, e_{n+1}$ arranged in increasing order.
We evaluate $\rho_n = (e_{n+1}-e_n)/(e_n-{e_{n-1}})$, and compute $\overline{\rho}$,
\begin{align}
\overline{\rho} = \overline{\min(\rho_n, 1/\rho_n)}. \label{eq:level_spacing_ratio} 
\end{align}
The Hamiltonians can be therefore classified in terms of their ratios: for a GOE Hamiltonian, like $H_{GG}$, $\overline{\rho} \approx 0.53$; for a Gaussian unitary ensemble (GUE) Hamiltonian $\overline{\rho}\approx 0.6$; lastly, for a Poisson Hamiltonian, like $H_{GP}$, one gets $\overline{\rho}_n= 2\ln2 -1 \approx 0.386$ \cite{OganesyanHuse2007}. 

\section{Results} \label{sec:results}
In this section, we present the results from studying the projected measurements and the corresponding correlation matrices.
Unless specified otherwise, we work with system sizes $L=10$.
We evolve the system using a matrix product states algorithm with sufficient bond dimension for exact evolution and use the approach described in Fig.~\ref{fig:zipperSetup} to extract the configurations. 

\begin{figure}[h]
    \includegraphics[width = \columnwidth]{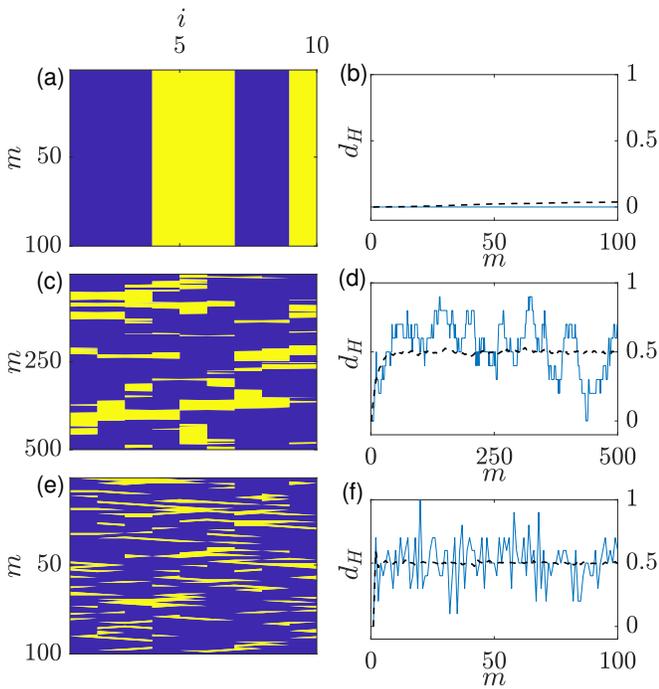}
		\caption{Panels (a, c, e): spin configuration measurements over time for the non-integrable Heisenberg model $H_{H}$ with $L = 10$. Darker (blue)
regions are for spins projected as 0; brighter (yellow) regions are for spins projected as 1. Panels (b, d, f): corresponding measurement of the Hamming distance, $d_H(1,m+1)$ between the first configuration and measurement step $m+1$, both for a single trajectory (continuous blue line), and averaged over $250$ trajectories (black dashed line). For all panels $J_x = J_y = 0.2, J_z = 1$, $h_x = 2$. Panels (a,b) are for $\Delta t=0.01$, (c,d) are for $\Delta t=0.1$, and (e,f) are for $\Delta t=1$.}
		\label{fig:oneTrajPlusHammingDist} 
\end{figure}

\subsection{Hamming distance}
In Fig.\ref{fig:oneTrajPlusHammingDist} we give an intuition on the dynamics one can expect from the measured spin configurations. 
Here we consider a non-integrable Heisenberg model $H_{H}$ in the paramagnetic region of the parameters space with, from Eq.~(\ref{Hamiltonian}), $J_x = J_y = 0.2, J_z=1$, and $h_x=2$. 
In panels (a, c, e) we show the evolution of the probabilistic cellular automata, with the rows being the configurations and the columns being the measurements. 
The darker regions are for spins projected as $0$ and the brighter regions are for spins projected as $1$.
In panels (b, d, f), we show the corresponding evolution of the Hamming distance $d_H(1,m+1)$ computed for a single trajectory, (dashed black line) and averaged over $250$ trajectories (continuous blue line).
In panels (a,b), (c,d), and (e,f), we use $\Delta t=0.01$, $0.1$, and $1$ respectively. 
Due to the quantum Zeno effect, we observe much more rapid dynamics in the probabilistic cellular automata in between measurements for larger $\Delta t$.
This is particularly evident from the Hamming distance averaged over many trajectories. 
In fact, this can reach values close to $0.5$, implying that, on average, half of the sites change their state from $0$ to $1$ or vice-versa.

\begin{figure}[h]
    \includegraphics[width = \columnwidth]{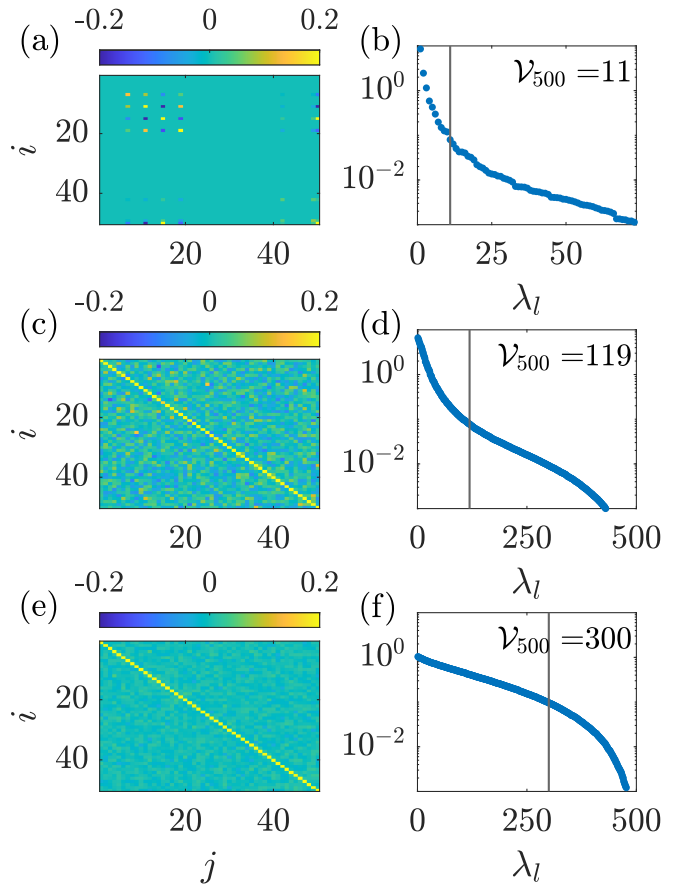}
		\caption{Panels (a, c, e): heat map of the correlation matrices $C^m_{i,j}$ at measurement step $m=500$. Panels (b, d, f): Spectrum of the eigenvalues of the correlation matrices respectively in panels (a, c, e). The trace of the eigenvalue spectrum gives the PCA complexity at a given time $m$, i.e. $\mathcal{V}_m$. The vertical lines indicate the $k$-th eigenvalue at which $\Tr(\lambda_{l<k})/\Tr(C) = 0.95$. In all panels we considered an $H_{H}$ model with non-zero parameters  $J_x = J_y = 0.2; J_z = 1; h_x = 2$. Note: panel (b) zooms into the first 75 eigenvalues since the remaining eigenvalues are negligible in comparison. Panels (a,b) are for $\Delta t = 0.01$, (c,d) are for $\Delta t = 0.1$ and (e,f) are for $\Delta t = 1$. }
		\label{fig:correlationMatAndDiags}
\end{figure}

\subsection{Evolution of principal component analysis}
In order to gain a deeper insight into the dynamics, we perform a PCA on the correlation matrix $C_{ij}^m$ in Eq.(\ref{eq:Cij}). 
In Fig.~.\ref{fig:correlationMatAndDiags}, we have considered the $H_{H}$ Hamiltonian with non-zero parameters $J_x = J_y = 0.2$, $ J_z = 1$, $ h_x = 2$.
We show the covariance matrix $C_{ij}^m$ in panels (a, c, e) and its eigenvalues in panels (b, d, f).
For all panels, $m=500$.
In panels (a,b), (c,d), and (e,f), we use $\Delta t=0.01$, $0.1$, and $1$ respectively.
Similarly to Fig.~\ref{fig:oneTrajPlusHammingDist}, we observe minimal evolution for small $\Delta t$.
For larger $\Delta t$, the correlation matrix starts to approach a rescaled identity matrix at long times, and the principal component analysis gives a large $\PCA_{500}$.
Qualitatively similar results can be reached for other Hamiltonians (not shown).

\begin{figure}
\includegraphics[width=\columnwidth]{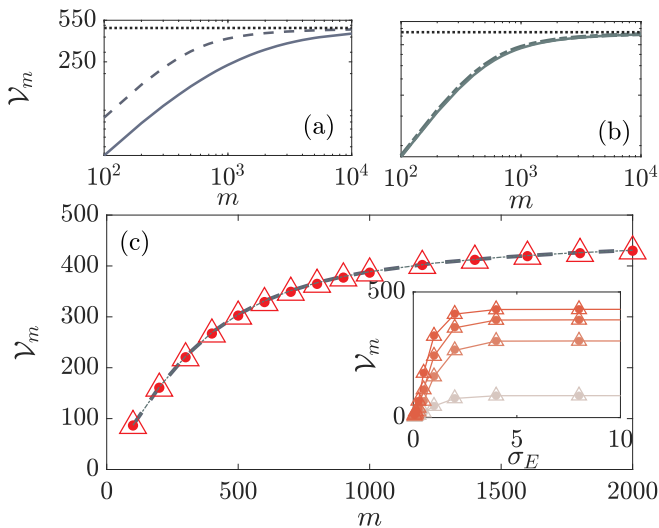}

\caption{Evolution of the principal component analysis value $\PCA_m$ using the (a) Ising and (b) $H_{H}$ Hamiltonians. The black dotted lines (a-b) show the maximum value obtainable by the principal component analysis: $\PCA_m = 475$. Panel (c) presents a comparison of random all-to-all Hamiltonians of type $H_{GG}$ (circle markers) and $H_{GP}$ (triangular markers), as well to Ising (dashed) and $H_{H}$ (dot-dashed). In panels (a-b), the blue dashed lines correspond to $h_x = 2$, while the continuous line corresponds to $h_x = 4$, and $J_z=1$, while the other non-zero parameters of the Hamiltonian $H_{H}$ in panel (b) are  $J_x = J_y = 2$ and $h_z = 0.5$. In all panels $\Delta t = 0.4$. The inset shows $\PCA_m$ versus the standard deviation of its spectrum at different measurement steps $m=100,\;500,\;1000$ and $2000$, from the lighter to the darker. The filled circles represent the data for $H_{GG}$, while the open triangles for $H_{GP}$ while the lines are guides to the eye.
} 
\label{fig:fig4}
\end{figure} 

In Fig.~\ref{fig:fig4}, we plot the evolution of the principal component analysis complexity $\PCA_m$ against the number of measurements $m$. 
We compare two different Hamiltonians, $H_I$ in Fig.~\ref{fig:fig4}(a) and $H_{H}$ in Fig.~\ref{fig:fig4}(b).
The former is an integrable non-interacting model and the latter is a non-integrable interacting model.  
In both Fig.~\ref{fig:fig4}(a) and (b), the dashed line corresponds to $h_x = 2$, the continuous line corresponds to $h_x = 4$, while $J_z$ is set to $1$.
we observe that the dynamics have two regimes: $\PCA_m$ first grows rapidly, then slowly saturates toward a finite value.
The quantitative character of the growth, though, is dependent on the model studied.
For $H_I$ in Fig.~\ref{fig:fig4}(a), increasing $h_{x}$ from $2$ to $4$ results in a substantial slower $\PCA_m$ growth.
However, for $H_{H}$ in Fig.~\ref{fig:fig4}(b), the same change in $h_{x}$ has little influence in the dynamics of $\PCA_m$.
In order to understand the differences, we repeat the same analysis for $H_{GG}$ (filled circles) and $H_{GP}$ (empty triangles) in Fig.~\ref{fig:fig4}(c). For this evaluation, once we have produced the all-to-all random Hamiltonians we have rescaled their energies so that the standard deviation of the spectrum would be the same as the Ising model $H_I$ with $h_x=2$. 
We find that both Hamiltonians produce the same $\PCA_m$ dynamics despite having different level spacing statistics $\bar{\rho}$.
This suggests that the difference in $\PCA_m$ dynamics is not tightly due to different level spacing statistics. Interestingly, in panel (c) we also plot the evolution of $\PCA_m$ from $H_I$ and $H_H$ with $h_x=4$ (respectively dashed and dot-dashed lines). We observe a clear match, indicating that the PCA complexity in a local system can match that of all-to-all Hamiltonians. To better understand this we study more carefully the role of the energy scale of the all-to-all Hamiltonians. In the inset of panel (c) we show the value of $\PCA_m$ for different measurement times $m$ versus the standard deviation of the energy of the Hamiltonians. A small standard deviation implies limited dynamics, while larger standard deviations result (at least up to the energy scales studied) in more consistent results, i.e. a flatter curve. The measurement steps considered are $m=100,\;500,\;1000$ and $2000$, from the lighter to the darker, while we still use triangles for $H_{GP}$ and filled circles for $H_{GG}$.   

\begin{figure}[h]
\includegraphics[width=\columnwidth, left]{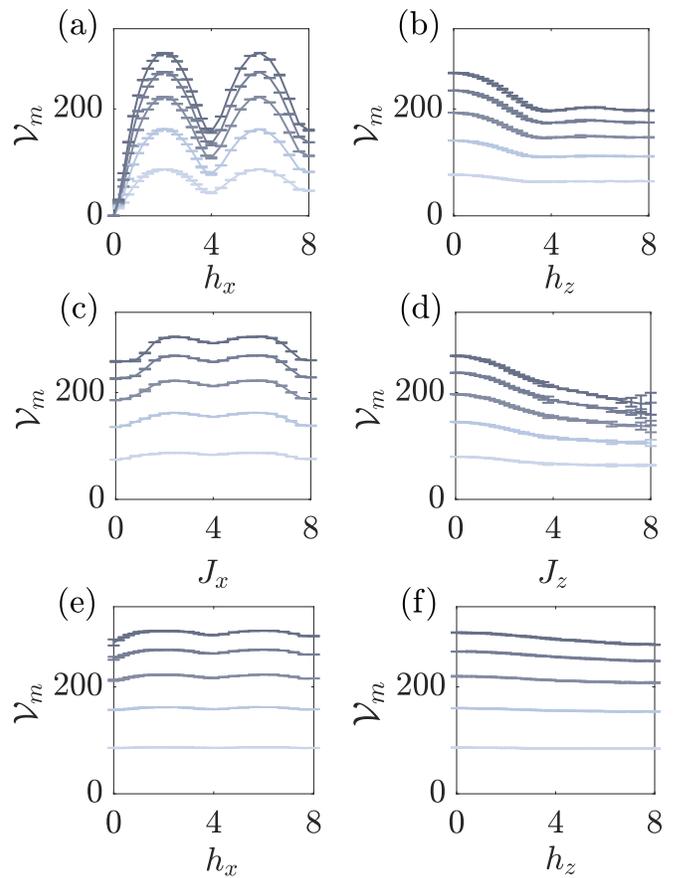}
\caption{PCA complexity versus one Hamiltonian parameter for fixed measurement interval $\dt = 0.4$, system size $L = 10$, $N=50$ and each point is averaged over 5 different realizations. All measurements were realized along the z-axis. Lighter colors (bottom lines) correspond to total time $m = 100$; darker lines correspond to increments of $m$ of $100$ steps. Panels (a) and (b) show results for the Ising model $H_I$ ($J_z = 1$ and $J_x=0$), and the rotated Ising model $H_{IR}$ ($J_x = 1$ and $J_z = 0$) respectively; Panels (c) and (d) the XYZ ($J_y = 1;~ J_z = 1.3$), and rotated XYZ ($J_x = 0.5;~ J_y = 1$) models respectively. Panels (e) and (f) the non-integrable Heisenberg model $H_{H}$ ($J_x = J_y = 2;~ J_z = 1;~h_z = 0.5$), and the rotated one $H_{HR}$ ($J_y = J_z = 2;~ J_x = 1;~ h_x = 0.5$). }
\label{fig:differentHamiltoniansPanel}
\end{figure}

We now investigate further the dynamics of the PCA complexity for local Hamiltonians.  
In Fig.~\ref{fig:differentHamiltoniansPanel}, we study the evolution of $\PCA_m$ for different types of Hamiltonians and for different numbers of measurement steps.
In each panel, the curves from bottom to top represent $\PCA_m$ for $m=100,\;200,\;300,\;400$ and $500$ respectively (from lighter to darker). Furthermore, each line is the average of 5 different realizations and the error bars show the standard deviation. 
In Fig.~\ref{fig:differentHamiltoniansPanel}(a) we consider the Ising model $H_I$ and vary $h_x$ from $0$ to $8$. 
For $h_x=0$, there is no growth of $\PCA_m$ because the dynamics are completely frozen on the computational basis. 
As $h_x$ increases, we observe strong oscillations in $\PCA_m$.
Our understanding is that this strong dependence on $h_x$ is due to both the non-interacting nature of this model and to the orientation of the measurements. 
In fact, after each measurement, the local $h_x$ field tends to rotate the spins.
To confirm this, we consider the rotated Ising model $H_{IR}$, in which the interaction in the $z$ direction is changed into the $x$ direction, but the measure is still done in the $z$ direction. 
In this case, at $h_z=0$ there is already a significantly large value of $\PCA_0$ because of the $\sigma^x_n\sigma^x_{n+1}$ terms in the Hamiltonian.
Increasing the value of $h_z$ results in slower $\PCA_m$ dynamics at all $m$, which becomes almost constant for large $h_z$.
The strong oscillation disappears because the local field and measurements are in the same direction.
We have thus shown that the direction in which the measurement occurs plays a quantitative role in the evolution of $\PCA_m$.
Comparing Fig.~\ref{fig:differentHamiltoniansPanel}(a) and (b), we also observe that, for the range of $m$ explored, there can be a significant difference in the magnitude of $\PCA_m$ for different values of the local field.
Since the Hamiltonians in Fig.~\ref{fig:differentHamiltoniansPanel}(a) and (b) are both integrable and non-interacting, we further consider the integrable and interacting model in Fig.~\ref{fig:differentHamiltoniansPanel}(c) and (d). 
We evaluate the XYZ Hamiltonian $H_{XYZ}$ in Fig.~\ref{fig:differentHamiltoniansPanel}(c), and the rotated XYZ Hamiltonian $H_{XYZR}$ in Fig.~\ref{fig:differentHamiltoniansPanel}(d).
Comparing the two Hamiltonians, we vary the interaction strength $J_x$ in $H_{XYZR}$ and $J_z$ in $H_{XYZ}$. 
The other terms are constant. 
We observe a similar behavior to what was observed in Fig.~\ref{fig:differentHamiltoniansPanel}(a) and (b). 
The main differences between Fig.~\ref{fig:differentHamiltoniansPanel}(a) and (c, d) are that: 
\begin{enumerate}
    \item $\PCA_m$ is not zero for $J_z$ or $J_x=0$ because $J_y$ is non-zero and thus allows the spins to change.
    \item The magnitude of the oscillation of $\PCA_m$ is quantitatively smaller in the interacting Hamiltonian.
\end{enumerate}
Fig.~\ref{fig:differentHamiltoniansPanel}(b) and (d) also share qualitatively similar conclusions regarding the role of the direction of the measurement, i.e., 
when the local field or interaction is in the same direction as the measurement, there is a clear decrease in $\PCA_m$ at all $m$ for an increase of this field or interaction.
We then consider the $\PCA_m$ dynamics for the $H_{H}$ Hamiltonian, where we vary $h_x$ in Fig.~\ref{fig:differentHamiltoniansPanel}(e) and $h_z$ Fig.~\ref{fig:differentHamiltoniansPanel}(f). 
A remarkable feature in the former is that the oscillations in the magnitude of $\PCA_m$ as a function of the Hamiltonian parameters are even smaller.
On the other hand, in Fig.~\ref{fig:differentHamiltoniansPanel}(f), we can still see a decrease of the $\PCA_m$ as we increase $h_z$. However, the variability of the $\PCA_m$ is greatly reduced when compared to the Hamiltonians presented in Fig.~\ref{fig:differentHamiltoniansPanel}(b) and (d). It does appear that the PCA complexity tends to be large and much less dependent on Hamiltonian and protocol parameters for the non-integrable model.  

\begin{figure}[ht!]
    \includegraphics[width=\columnwidth]{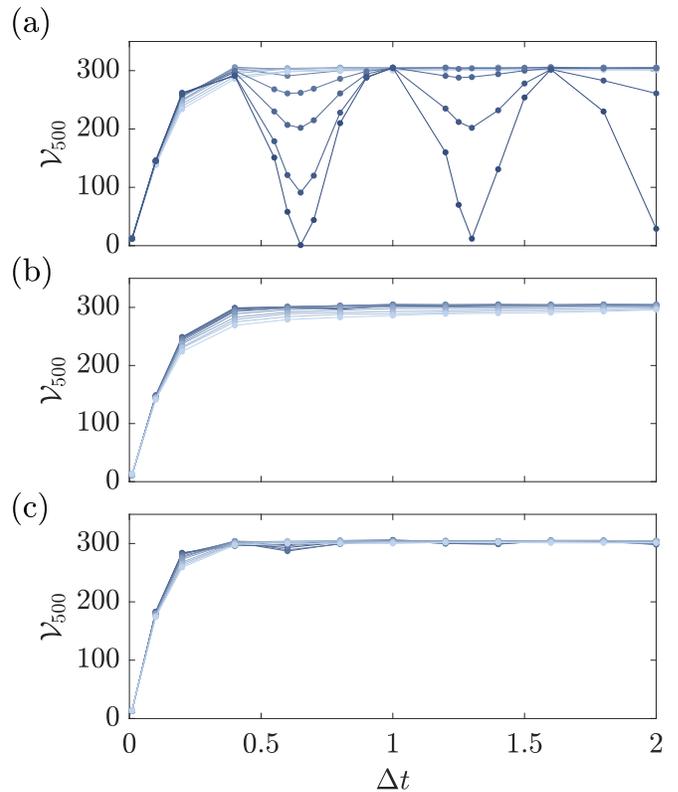}
    \caption{PCA complexity after $500$ measurement versus measurement time $\Delta t$. Panels (a-c) are for the Ising model $H_I$, the XYZ model $H_{XYZ}$, and the non-integrable Heisenberg model $H_{H}$ respectively. In all panels, the tunneling $J_z$ goes from zero to $4$, in $0.4$ increments (from darker to lighter). Additionally, for panel (a) $h_x = 2.4$; In panel (b), $J_x = 1, J_y = 1.7$; in panel (c) $J_x = J_y = 1,~h_x = 2.4$, 
and $J_z$ goes from zero to $4$, in $0.4$ increments (from darker to lighter). For all cases $L = 10$ and $N=50$.}
    \label{fig:differentMeasurementTimes}
\end{figure}

We now evaluate in more detail the dependence of $\PCA_m$ on the measurement interval $\Delta t$. 
In Fig.~\ref{fig:differentMeasurementTimes}, we consider the same three Hamiltonian types presented in Fig.~\ref{fig:differentHamiltoniansPanel}, the Ising model $H_I$, Fig.~\ref{fig:differentMeasurementTimes}(a), the XYZ model $H_{XYZ}$, Fig.~\ref{fig:differentMeasurementTimes}(b), and the non-integrable Heisenberg model $H_{H}$, Fig.~\ref{fig:differentMeasurementTimes}(c). 
In all panels, $\PCA_m$ for $m=500$ is plotted versus $\Delta t$.
Every line corresponds to different values of the Hamiltonian parameter $J_z$.
When $\Delta t$ is small, the dynamics are particularly slow for all Hamiltonians due to quantum Zeno dynamics.
The most remarkable difference, when comparing Fig.~\ref{fig:differentMeasurementTimes}(a) and (b, c), is that the variance in $\PCA_m$ is significantly smaller in (b, c) than in (a).
Recall that such oscillations in $\PCA_m$ of $H_I$ are similarly observed in Fig.~\ref{fig:differentHamiltoniansPanel}(a).
The interplay between measurements and local dynamics results in a resonant interplay occurring for $h_x \Delta t = p\pi/2$ where $p$ is an integer.  
This suggests that, while the particular quantitative details of the dynamics depend on the properties of the Hamiltonian studied, more general behaviors (like the variance of the oscillations versus $\Delta t$ or other Hamiltonian parameters) depend on how much of the dynamics are influenced by the single-site rotations perpendicular to the measurement direction. 
This can be significant in a non-interacting Hamiltonian.
However, if the Hamiltonian is interacting, more so if it is also non-integrable, the effective single-site dynamics are suppressed and the oscillation in $\PCA_m$ disappears. Hence, this analysis can be seen as a diagnostic tool to detect emergent non-interacting dynamics. 

\begin{figure}[h]    \includegraphics[width=\columnwidth]{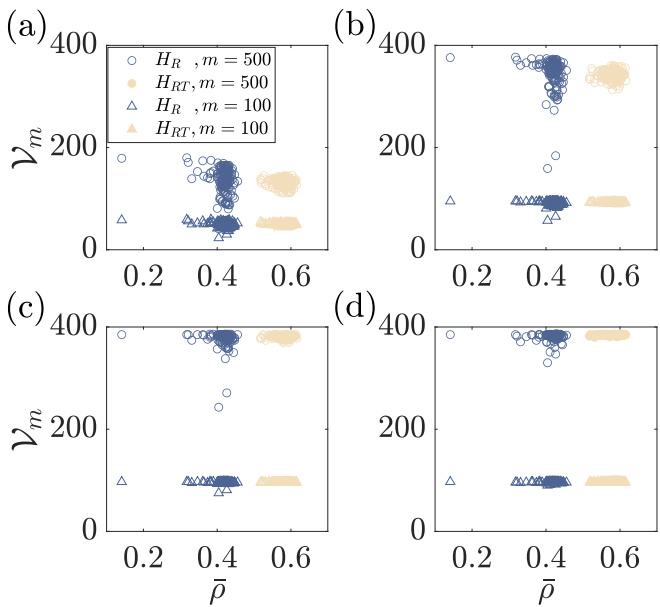}
    \caption{PCA complexity $\PCA_m$ versus level spacing statistics ratio $\bar{\rho}$ for local random Hamiltonians without, $H_R$, and with, $H_T$, translational invariance: $H_R$ in dark blue and $H_{RT}$ is light brown at measurement steps $m=100$ (triangles) and $m=500$ (circles). The panels (a-d) depict the scenarios $\Delta t = 0.01$, 0.05, 0.1, and 0.2 respectively. } \label{fig:PoissonVsGUEoverR}
\end{figure} 

\subsection{Principal component analysis, random local Hamiltonians, level statistics and pinning}
To better understand the main contributors to the evolution of the PCA complexity $\PCA_m$, we now consider the two random local Hamiltonians $H_{R}$ and $H_{RT}$ from Eq.(\ref{Hamiltonian}). 
The former has Hamiltonian parameters that are random (drawn from a normal distribution with zero mean and unit standard deviation) for all sites but the latter has parameters that are random and translationally invariant.  
We stress that, in order to have fair comparisons between the random Hamiltonians, once we produce a random Hamiltonian, we rescale the Hamiltonians such that all of them have the same standard deviation in the energies.
For each type of Hamiltonian, we consider $150$ different sets of random Hamiltonians. 

In Fig.~\ref{fig:PoissonVsGUEoverR}, we plot the values of $\PCA_m$ against the level spacing statistics $\overline{\rho}$ of the different Hamiltonians used.
The choice of $\Delta t$ for (a, b, c, d) is $0.01, 0.05, 0.1, 0.2$ respectively.
In all panels, we use dark blue symbols for $H_R$, and light brown for $H_{RT}$. 
We use triangles for measurement steps $m=100$ and circles for $m=500$. 
We observe clustering of the $\PCA_m$ for $H_R$ around $\overline{\rho}\approx 0.386$, which is the value for Poisson statistics.
This may be due to the local disordered nature of the Hamiltonian. 
One the other hand, the random and translationally invariant $H_{RT}$ has larger values of $\overline{\rho}$ between $0.53$ (GOE) and $0.6$ (GUE), 
$H_{RT}$ can thus be considered as chaotic.
When compared against $H_R$, we notice that $H_{RT}$ has a smaller variance in the spread of $\PCA_m$, both at $m=100$ and $m=500$. 
This is verified for all the values of $\Delta t = 0.01, 0.05, 0.1, 0.2$ tested in Fig.~\ref{fig:PoissonVsGUEoverR}.

\begin{figure}[h]
    \includegraphics[width=\columnwidth]{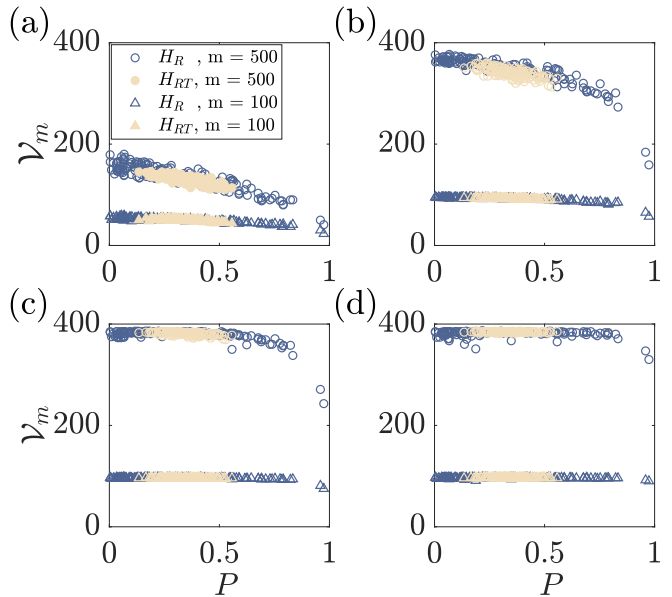}
    \caption{PCA complexity $\PCA_m$ versus pinning $P$ for two different types of Hamiltonians: $H_R$ in dark blue and $H_{RT}$ is light brown at measurement steps $m=100$ (triangles) and $m=500$ (circles). The panels (a-d) depict the scenarios $\dt = 0.01$, 0.05, 0.1 and 0.2 respectively.}
    \label{fig:PoissonVsGUEoverPining}
\end{figure}
To gain further insight, we consider the same data but plot them against the pinning coefficient $P$, which we define as
\begin{equation} 
    P =  \dfrac{\sum_i (|h_{z,i}|^2 + |J_{z,i}|^2)}{\sum_{j,a=x,y,z}(|h_{a,i}|^2 + |J_{a,i}|^2)}.   
\end{equation}
The idea of this coefficient is that a large value would imply that the terms in the Hamiltonian are dominated by the operators $\sigma^z_i$ and $\sigma^z_i \sigma^z_{i+1}$, which lead to no change in the configuration after a measurement in the computational basis. 
Hence, for $P\approx 1$ we would expect much slower dynamics, especially for interacting and non-integrable systems that suppress single particle dynamics.
Indeed we observe this in Fig.~\ref{fig:PoissonVsGUEoverPining}.
The choice of $\Delta t$ for (a, b, c, d) is $0.01, 0.05, 0.1, 0.2$ respectively.
In all panels, we use dark blue symbols for $H_R$, and light brown for $H_{RT}$. 
We choose the triangles for measurement steps $m=100$ and circles for $m=500$. 
While the $\PCA_m$ for $H_{RT}$ are closer to each other, and correspond to both a smaller pinning $P$ and a smaller range of pinning values, the random local Hamiltonians $H_R$ result in a much larger spread of $\PCA_m$ and pinnings $P$. 
It is for these large values of $P$, which occur only for $H_R$, that we observe much slower dynamics, denoted by a smaller value of $\PCA_m$.  

\section{Conclusions} \label{sec:conclusions}
We have studied a quantum process in which unitary evolution is followed by projective measurements. 
We have observed that the dynamics depend on both the Hamiltonian used and the measurement intervals. 
In particular, larger time intervals $\Delta t$ correspond to a faster increase of the Hamming distance between consecutive measurements.
We have focused on the complexity of the sequences of spin configurations produced with this evolution protocol.
To do so, we have used the principal component analysis on the correlation matrices generated at different times. 
We have considered different known models, such as the Ising model $H_I$ as an example of an integrable non-interacting Hamiltonian, the XYZ model $H_{XYZ}$ as an example of interacting integrable Hamiltonian, and an XXZ model plus transverse and longitudinal fields $H_{H}$ as an example of interacting non-integrable Hamiltonian. 

We found that the PCA complexity grows rapidly before plateauing, and this general behavior is observed in all the scenarios considered. Interestingly, the qualitative behavior can be significantly affected by the Hamiltonian parameters used, and even more importantly, local Hamiltonians can result in an evolution of the PCA complexity which matches that of global Hamiltonians, both chaotic or not. 

We have thus observed that $\PCA_m$ may oscillate significantly when changing the Hamiltonian parameters or the measurement intervals, but much less so in the case of interacting and/or non-integrable models.
We have also considered a set of random Hamiltonians, both with and without translational invariance.
Those without translational invariance generally follow Poisson statistics and they show PCA complexities $\PCA_m$ with a broader set of values when compared to translationally invariant random Hamiltonians that tend to follow GOE/GUE statistics.       
Lastly, when studying the $\PCA_m$ data against the pinning coefficients we observed that most random translationally invariant Hamiltonians have lower pinning and larger complexity.
For non-translationally invariant Hamiltonians $H_{R}$, we obtain a broad spectrum of the PCA complexity, and $\PCA_m$ becomes clearly smaller for pinning $P$ approaching one. 
It is because the translationally invariant random Hamiltonians $H_{RT}$ are unlikely to have large pinning values that they have a much smaller variance of $\PCA_m$ for different sets of Hamiltonian parameters.    

The dynamics studied in this work could be readily tested in state-of-the-art experiments including Rydberg atoms with optical tweezers \cite{BrowaeysLahaye, SchollBrowaeys2022, Lukin, Loh},  ultracold atoms with a microscope \cite{BakrGreiner2009, GrossBloch2017, GrossBakr2021}, ion traps \cite{BlattRoos2012,MonroeKim2013}, arrays of superconducting qubits \cite{DevoretMartinis2004, KjaergaardOliver2020}. 
Future studies could also focus on a quantitative study of the short-time algebraic growth of the principal component analysis as a function of the Hamiltonian considered and measurement protocols. In addition, some of the diagnostics introduced here can readily be adapted to investigate the dynamics of circuits, both in the context of concrete quantum algorithms and in that of non-energy conserving time evolution of physical systems. 

\begin{acknowledgments}
D.P. acknowledges support from the Ministry of Education Singapore, under the grant MOE-T2EP50120-0019. D.P. also acknowledges discussions with Vint Ve and Zhao Yu Ma. The computational work for this article was partially performed at the National Supercomputing Centre, Singapore \cite{NSCC}. 
H.P.C acknowledges the important input from Gabriel T. Landi, Xiansong Xu, Rafael M. Magaldi, and Stephane Bressan, as well as the help from Douglas Casagrande with some of the simulations. Simulations were performed using the ITensor library ~\cite{ITensor, oDMRG}. The codes and data generated are both available upon reasonable request to the authors.

\end{acknowledgments}

\normalem
\bibliographystyle{apsrev4-1}
\bibliography{refs} 

\end{document}